\documentclass[twocolumn]{aastex62}

\usepackage{color}
\usepackage{wasysym}

\newcommand{\mytilde}{\raise.17ex\hbox{$\scriptstyle\sim$}}
\newcommand{\brk}[1]{$[$#1$]$}

\newcommand{\kepler}{{\it Kepler}}
\newcommand{\ktwo}{{\it K2}}

\newcommand{\tess}{{\it TESS}}
\newcommand{\hst}{{\it HST}}
\newcommand{\spitzer}{{\it Spitzer}}

\newcommand{\mps}{m~s$^{-1}$}

\newcommand{\Rear}{${R_\oplus}$}

\newcommand{\Teff}{${T_{\rm eff}}$}

\newcommand{\ntran}{$N_{\rm tran}$}

\newcommand{\Logg}{$\log{\rm g}$}

\newcommand{\Porb}{$P_{\rm orb}$}

\newcommand{\Rp}{$R_{\rm p}$}
\newcommand{\Kp}{$K_{\rm p}$}
\newcommand{\ntarg}{$N_{\rm t}$}
\newcommand{\ncand}{$N_{\rm c}$}
\newcommand{\nmtarg}{$N_{\rm m}$}
\newcommand{\nktarg}{$N_{\rm k}$}
\newcommand{\nthreemcand}{$N_{\rm 3+c}$}

\newcommand{\Pgd}{$P$}
\newcommand{\mboost}{$\pi_{\gamma_m}$}
\newcommand{\cjb}{}
\newcommand{\cjbtwo}{}
\newcommand{\cjbthree}{}

\usepackage[normalem]{ulem}

\usepackage{xcolor, fontawesome}
\definecolor{twitterblue}{RGB}{64,153,255}
\newcommand{\twitter}[1]{\href{https://twitter.com/#1}{\textcolor{twitterblue}{\faTwitter}\,\tt \textcolor{twitterblue}{@#1}}}

\accepted{December 26, 2018 - The Astronomical Journal}

\begin{document}
\title{RE-EVALUATING SMALL LONG-PERIOD CONFIRMED PLANETS FROM {\it KEPLER}}

\author[0000-0002-7754-9486]{Christopher~J.~Burke}
\affiliation{Kavli Institute for Astrophysics and Space Research, Massachusetts Institute of Technology, Cambridge, MA \twitter{CuriousTerran}}

\author{F.~Mullally}
\affiliation{SETI Institute, Mountain View, CA \twitter{fergalm}}

\author[0000-0001-7106-4683]{Susan~E.~Thompson}
\affiliation{Space Telescope Science Institute, Baltimore, MD \twitter{mustaric}}

\author[0000-0003-1634-9672]{Jeffrey~L.~Coughlin}
\affiliation{SETI Institute, Mountain View, CA \twitter{JeffLCoughlin}}

\author[0000-0002-5904-1865]{Jason~F.~Rowe}
\affiliation{Department of Physics and Astronomy, Bishop's University, Sherbrooke, Canada \twitter{JasonFRowe}}

\begin{abstract}

We re-examine the statistical confirmation of small
long-period \kepler\ planet candidates in light of recent improvements
in our understanding of the occurrence of systematic false alarms in
this regime. Using the final Data Release 25 (DR25) Kepler planet candidate catalog statistics, we find that the previously confirmed single
  planet system Kepler-452b no longer achieves a 99\% confidence in
  the planetary hypothesis and is not considered statistically
  validated in agreement with the finding of \citet{MUL18}.  For multiple planet
  systems, we find that the planet prior enhancement for belonging to
  a multiple planet system is suppressed relative to previous
  \kepler\ catalogs,
and we identify the multi-planet system member,
  Kepler-186f, no longer achieves a 99\% confidence in the planetary
  hypothesis.
{\cjbthree Because of the numerous confounding factors in the data analysis process that leads to the detection and characterization of a signal, it is difficult to determine whether any one planetary candidate achieves a strict criterion for confirmation relative to systematic false alarms. For instance, when taking into account a simplified model of processing variations, the additional single planet systems Kepler-443b,
Kepler-441b, Kepler-1633b, Kepler-1178b, and Kepler-1653b have a non-negligible probability of falling below a 99\% confidence in the planetary hypothesis.}
The systematic false alarm hypothesis must be taken into account when
employing statistical validation techniques in order to confirm planet
candidates that approach the detection threshold of a survey.  We
encourage those performing transit searches of \ktwo, \tess, and other
similar data sets to quantify their systematic false alarms rates. Alternatively, independent photometric detection of the transit signal
or radial velocity measurements can eliminate the false alarm
hypothesis.

\end{abstract}

\keywords{methods: statistical --- planetary systems --- planets and satellites: detection --- planets and satellites: terrestrial planets --- surveys --- techniques: photometric}

\section{Introduction}\label{sec:intro}

The final \kepler\ DR\,25 pipeline run \citep{TWI16} and
\kepler\ science office Robovetter classification \citep{THO17}
provides an unprecedented uniform catalog of planet candidates for
understanding the dynamics and occurrence outcomes of the planet
formation process.  The planet candidate catalog was accompanied by an
equally impressive set of supplemental data to help measure the biases
and corrections necessary to make full use of the catalog.  The following is a
non-exhaustive list of supplemental data products associated with the
DR\,25 \kepler\ data
\begin{itemize}
\item Stellar parameter catalog \citep{MAT17}
\item Database of transit signal injection followed by the full pipeline and classification recovery \citep{CHR17,COU17A}
\item Database of transit signal injection followed by the transiting planet search module alone for greater detail into signal recovery \citep{BUR17A,BUR17B}
\item \kepler\ pipeline open software release\footnote{https://github.com/nasa/kepler-pipeline} and documentation \citep{JEN17} for understanding implementation details of the pipeline algorithms
\item Astrophysical false positive classification \citep{MOR16} as well as false positive analysis taking into account ground-based follow-up observations \citep{BRY17}
\item False positive classification taking into account the stellar image position \citep{MUL17}
\item  Characterization of the systematic false alarms through data permutation techniques \citep{COU17A,THO17}
\end{itemize}
The last item offers new insight into quantifying the Type I error
rate of a transit survey.  \citet{JEN02} used theoretical arguments to
propose that a 7.1\,$\sigma$ threshold would produce $<$ 1 false alarm
(or Type I error) in the entire \kepler\ data set under certain
assumptions regarding the expected noise.  However, experimental data,
even after detrending, can contain residual systematics in excess of
the expected noise causing an enhanced false alarm rate.  The
simulated false alarms described in \citet{COU17A} represent the first
time a transit survey has quantified its Type I false alarm rate based
upon the experimental data for the express purpose of constraining the
detections induced by instrumental or data processing systematics that
can mimic a transit signal.  It was recognized in analyzing the Q1-Q16
\kepler\ planet candidate catalog \citep{MUL16} that there was
evidence for excess detections at long periods, likely due to
low-level systematic false alarms that mimic the low signal-to-noise
ratio, few transit events highly sought after for detection.
\citet{BUR15} concluded that quantifying the rate of the false alarm
contamination was a leading systematic uncertainty in constraining the
GK dwarf habitable zone planet occurrence rate.

Astrophysical false positives, such as background eclipsing binary
signals, were also identified in the nascent stage of transit
discoveries as a contaminating source of false positive transit
signals \citep{BRO03}.  Thus, transit discoveries relied heavily on
radial velocity confirmation and follow-up observations to vet against
astrophysical sources of contamination \citep{TOR04,MAN05}.
Systematic false alarms in the discovery data (typically based upon
small telescopes) could be rejected by additional photometric
follow-up at the predicted transit times (typically using larger
telescopes) in order to provide an independent detection of the
transit signal and confirm its ephemeris.  While the high signal-to-noise ratio (SNR) of most candidates eliminates the concern for
systematic false alarms, low SNR detections from \kepler\ are
extremely difficult to follow-up photometrically from the ground due
to their long orbital periods, long transit durations, and faint host
magnitudes.

Statistical validation methods were developed to vet against
astrophysical false positives for \kepler\ candidates
\citep{TOR11,FRE11,MOR11,LIS12,MOR12,DIA14,LIS14,ROW14}.  Although
statistical validation achieved success in constraining the
possibilities for astrophysical false positives, systematic false
alarms became a problem for \kepler\ candidates near the detection
threshold.  However, up to and including the Q1-Q12 \kepler\ planet
candidate catalog \citep{BOR11A,BOR11B,BAT13,BUR14,ROW15}, it was
possible to mitigate systematic false alarm contamination by examining
the extra 6-12 months of \kepler\ data that was available
post-discovery to ensure that the transit signal detected in less data
continued to increase in significance.  Starting with the Q1-Q16
\kepler\ catalog \citep{MUL16} and the end of the prime mission field for
observations, looking ahead at more data was no longer an option for
dealing with systematic false alarms.

Although the \kepler\ false alarm data set from \citet{COU17A} were
primarily focused on addressing the occurrence rate contamination problem,
\citet{MUL18} recently have demonstrated that the false alarm rate as
a contaminating source impacts the statistical validation method
employed to confirm planet candidates.  They conclude that the planet
Kepler-452b \citep[closest approximation to the planet radius, \Rp,
  and orbital period, \Porb, of Earth orbiting a Sun-like
  star;][]{JEN15} is not a confirmed planet anywhere close to the
99.7\% probability claimed by \citet{JEN15}.

In this paper, we follow on and expand the work by \citet{MUL18} to
address the full sample of \kepler\ confirmed planets in the
small-planet, long-period regime.
In Section~\ref{sec:reliab} we show how to frame the systematic false
alarm contamination problem in terms of the Bayesian probability
framework adopted by the statistical validation literature.  In
Section~\ref{sec:resultsing} we apply this framework to the full
sample of \kepler\ confirmed planets orbiting targets hosting a single
planet candidate, and Section~\ref{sec:multis} extends the
calculation to targets hosting multiple planet candidates.
Sections~\ref{sec:resultsing} and \ref{sec:multis} update the planet
probability in order to identify several previously confirmed planets
that can no longer be confirmed at the 99\% reliability level.  Section~\ref{sec:addanalysis} provides an alternative analysis that focuses on targets on the electronically `quiet' \kepler\ detectors.
Section~\ref{sec:discussion} discusses options for reconfirming these
newly unconfirmed planets and provides guidance for the \ktwo\ and
\tess\ missions \citep{HOW14,RIC16} for avoiding systematic false
alarms in their statistical validations.  Finally,
Section~\ref{sec:conclusion} summarizes the findings of this study.

\section{Confirming Planets Relative to Systematic False Alarms}\label{sec:reliab}

In this study, we follow the procedure for statistical validation of
transiting planet signals using a Bayesian framework \citep{TOR11,
  FRE11, MOR11, MOR12}.  In summary, the validation calculation
proceeds by specifying the likelihood that a signal of interest
matches what is expected of bona-fide planet ($\mathcal{L}_{{\rm
    PL}}$), astrophysical false positive ($\mathcal{L}_{{\rm AFP}}$),
and a systematic false alarm ($\mathcal{L}_{{\rm SFA}}$), as well as
a prior probability for each scenario ($\pi({\rm PL})$, $\pi({\rm
  AFP})$, $\pi({\rm SFA})$).  Overall, the Bayesian probability that a
\kepler\ signal originates from a planet given the data is,
\begin{equation}
  P({\rm PL}|{\rm data}) = \frac{\pi({\rm PL})\mathcal{L}_{\rm PL}}{\pi({\rm PL})\mathcal{L}_{\rm PL} + \pi({\rm AFP})\mathcal{L}_{\rm AFP} + \pi({\rm SFA})\mathcal{L}_{\rm SFA}}.
\end{equation}
For this study, we concentrate on the probability of the signal
originating from a planet relative to a systematic false alarm given
the data (i.e., $P({\rm PL}|{\rm data})/P({\rm SFA}|{\rm data})$) and
ignore the astrophysical false positive probability.  Ignoring the
astrophysical false positive probability is justified as we restrict
ourselves to the subset of \kepler\ discoveries that are considered
confirmed planets, where previous literature has validated the planet
relative to astrophysical false positives alone (i.e., $P({\rm
  AFP}|{\rm data})\, <\, 0.01 P({\rm PL}|{\rm data})$).  Thus, we need
to quantify $\pi({\rm PL})\mathcal{L}_{\rm PL}/\pi({\rm
  SFA})\mathcal{L}_{\rm SFA}$.

\subsection{Prior Probabilities and Likelihoods}

{\cjbtwo For the planet prior probability, $\pi({\rm PL})$, we adopt
  the observed \kepler\ objects of interest (KOIs) classified as
  planet candidates.  Using the \kepler\ planet candidate detections
  for measuring the planet prior probability was adopted by
  \citet{TOR15} to confirm numerous \kepler\ planets.  The
  \citet{TOR15} measurement of the planet prior consists of counting
  the number of \kepler\ planet candidates having \Porb\ within a
  factor of two of the candidate and \Rp\ within 3\,$\sigma$ of the
  measured planetary radius, and then scaling the resulting count by
  the number of targets surveyed.  This quantification of $\pi({\rm
    PL})$ assumes that astrophysical false positives (and systematic
  false alarms) are rare and that a majority of the planet candidates
  discovered by \kepler\ are bona-fide planets.  Thus, this estimate
  of $\pi({\rm PL})$ is an overestimate, but from the expected rates
  of astrophysical blend scenarios, it is accurate enough within a
  factor of a few for making statistical validations
  \citep{FRE11,MOR12}.  Rather than estimating the planet prior
  distribution through summing detections in a predefined region of
  parameter space, we employ kernel density estimation (KDE)
  techniques to estimate the planet prior as described below.}

In order to quantify $\pi({\rm SFA})$, it would be preferable to have
a quantitative model for the shapes, amplitudes, frequency, and
environmental drivers of the systematics present in the \kepler\ data.
Such a model is currently not available for \kepler, especially for
the level of SNR approaching the threshold for detection of transit
signals in the \kepler\ pipeline analysis MES=7.1
\citep{JEN02,CHR12,JEN17}, where MES (Multiple Event Statistic)
represents the SNR for detection in the \kepler\ pipeline.  MES is the
test statistic of the \kepler\ pipeline search algorithm which
quantifies the presence of a transit signal relative to the null
hypothesis.  As opposed to the hypothesis statistic for a single
transit event (Single Event Statistic, SES), the MES is measured over
multiple transit events that contribute to the signal.  Instead, in
order to measure $\pi({\rm SFA})$, we rely upon the \kepler\ data
itself in order to provide a model for false alarm signals.  The key
insight to quantifying $\pi({\rm SFA})$ is to modify the
\kepler\ relative flux time series in such a manner that the original
periodic dimming events of a transiting planet no longer coherently
combine to make detections.  We use data permutation techniques of
flux inversion and data scrambling (described below) in order to
suppress transit signals in the original data and generate modified
time series.  The modified time series can be re-searched for transit
signals, and the detected signals that are classified as `planet
candidates' are almost entirely due to systematic false alarms.
Re-analyzing the full, modified \kepler\ data set required automating
the classification process (i.e., the `Robovetter') as described in
\citet{MUL15}, \citet{COU16}, and \citet{THO17}.  Our definition of
$\pi({\rm SFA})$ is the same as the definition of $\pi({\rm PL})$, but
we take our `planet candidates' from the search and classification
performed on the flux inversion and data scrambling analysis and divide
by the number of targets contributing to detections.

\citet{COU17A} and \citet{THO17} describe our two methods for
generation of the simulated false alarms, but we summarize them here.
The first data modification, which we denote as inversion (INV), takes
advantage of the fact that the transit signals we seek have a flux
deficit.  The amplitude of the relative flux time series is modified
with a negative sign, such that dimming (brightening) signals become
brightening (dimming) signals, respectively.  As the planet search
algorithm returns dimming signals for consideration, the data
inversion converts dimming periodic transit signals into periodic
brightening signals, and the systematic false alarms (that previously
were periodic brightening events in the unmodified flux time series)
become detectable planet candidate signals by the search algorithm.
Data inversion relies on a key assumption that the systematic false
alarms contaminating the planet candidate signals are symmetric upon
data inversion.  While some false alarms are symmetric under
inversion, such as those due to `rolling-band' pattern noise
\citep{KOL10,CAL10}, other false alarms are not.  For example, the
cosmic ray induced sudden pixel sensitivity dropout
\citep[SPSD;][]{JEN10,STU12} produces a drop in flux followed by a
slow recovery.  Another weakness of data inversion is that there is
only one synthetic data set that can be generated by this process.

The second data modification employs statistical resampling of the
data employing a process similar to block bootstrap methods which we
denote as data scrambling (SCR).  In order to preserve the time
correlated structure (on time scales 1$<\tau<24$ hr) of the false
alarms that mimic transit signals while removing the repetitive
transit signal on \Porb\ time scales, we reorder data using yearly or
\kepler\ quarter blocks.  For instance, the normal time ordering of
\kepler\ observing quarters,
  [Q1, Q2, Q3, Q4, Q5, Q6, Q7, Q8, Q9, Q10, Q11, Q12, Q13, Q14, Q15, Q16, Q17],
were reordered into 
[Q13, Q14, Q15, Q16, Q9, Q10, Q11, Q12, Q5, Q6, Q7, Q8, Q9, Q1, Q2, Q3, Q4, Q17]
for one alternative permutation of the data.  The block data
scrambling/permutation tests performed using the DR\,25 pipeline
search are described in more detail in \citet{COU17A} and
\citet{THO17}.  The data scrambling test allows us to test the
hypothesis that systematic signals, randomly distributed in time,
align by chance and mimic a repeating transit signal, especially at
the low SNR, few number of transit (\ntran=3) limit of the
\kepler\ pipeline search.  By performing independent data reordering, we can
generate many synthetic data sets.  However, the time consuming nature of
performing the full pipeline and classification steps in the exact
same fashion as the observed/unpermuted data limited us to three
scrambled iterations.  One limitation of the block permutation is
astrophysical and transit signatures with periodicities shorter than
the block time scale remain coherent and can lead to detections
matching their presence in the original data.  Fortunately, this is
not an issue in the domain of interest in this paper.

Note that astrophysical phenomena can still generate detections in the
modified data sets.  Strong periodic signals \citep[deep eclipsing
  binaries, `heartbeat' tidal induced variables, and microlensing
  white dwarfs to name a few][]{SAH03,THO12,KRU14} can contaminate the
false alarm detections detected in the modified data sets.  Thus, the
modified planet candidate list is filtered \citep[see Section~2.3.3
  of][]{THO17} to avoid contamination by transiting planets, eclipsing
binaries, heartbeat stars, and self-lensing binary systems.  In
addition to the contamination list from \citet{THO17}, and since we
are interested in the false alarm contamination among the sample of
previously confirmed planets (which generally are of 'higher quality'
and undergo more inspection than the typical \kepler\ planet
candidate), we visually inspect the planet candidates from the
modified data sets to remove ones with transit signals that do not fit
the qualitative expectation for a repeatable transit signal.  This
by-eye filtering simulates the process by which planet candidates are
inspected to identify the most promising candidates suitable for
confirmation based upon how well they match the expectation for a
transit signal.  Although not ideal from a statistical and/or
repeatability aspect, visual inspection of the flux time series phased
to the orbital period of the detection is one of the most common
techniques employed in transit surveys for making decisions on what to
prioritize for observational and analysis follow-up.  Qualitatively,
one prioritizes transit signals that possess a clear box/u-shaped flux
decrement offset from the out-of-transit data.  In addition, `good'
transit signals have out-of-transit data that is well behaved and does
not show evidence for coherent signals of similar depth and duration
to the transit signal of interest.  From the inversion test, we remove
the planet candidates found for \kepler\ input catalog (KIC)
identifiers 5795353, 10782875, and 7870718.  From scrambling test two,
we remove planet candidates found for KIC 7531327, 10336674, and
10743850.  From scrambling test three, we remove planet candidates
found for KIC 5778913, 7694693, and 8176169.  No planet candidates
were removed from scrambling test one.  We do not perform the same
visual inspection of the planet candidates from the unmodified data,
thus our final sample likely underestimates $\pi({\rm SFA})$.

We simplify the problem further by assuming that $\mathcal{L}_{\rm
  PL}=\mathcal{L}_{\rm SFA}$ (i.e., a systematic false alarm signal and a
transiting planet signal have equal probabilities of explaining the
data).  This assumption is valid as the classified planet candidates
from the modified and unmodified data sets must pass a battery of 52
quantitative metrics that ensure that the signal is consistent with
the transiting planet hypothesis.  In particular, the Marshall test
\citep{MUL16} employed by the \kepler\ Robovetter, performs a formal
Bayesian model comparison test to ensure that a transit model has
comparable (or higher) probability for fitting the individual transit
events versus the other models considered such as a step discontinuity
or SPSD.  However, metrics like this are ineffective in the low SNR limit
where the detailed shape of the events can not be measured. Thus, our
detections are equally well modeled as a transit or a
systematic. Visual examination of the data \citep[such as from the
  \kepler\ data validation reports and \kepler\ science office vetting
  documents; ][]{COU17B} for the planet candidates found in the
modified data sets shows that they are qualitatively indistinguishable
from the planet candidates found in the unmodified data series
\citep[see ][for an example]{MUL18}.  With equivalent likelihoods, the
odds ratio probability $P({\rm PL}|{\rm data})/P({\rm SFA}|{\rm
  data})$ simplifies to the prior probability ratio, $\pi_{\rm
  r}=\pi_{\rm PL}/\pi_{\rm SFA}$.

\section{Results For Single Planet Systems}\label{sec:resultsing}

In order to calculate $\pi({\rm PL})$ and $\pi({\rm SFA})$, we use a
subset of the \kepler\ targets that are optimized for studying planet
occurrence rates for GK dwarfs.  See \citet{BUR15}, \citet{CHR17}, and
\citet{BUR17B} for a discussion, but in summary, in addition to
selecting targets based upon their spectral type, targets are selected
for their suitability as relatively quiet, well-behaved targets with
well modeled recoverability of transit signals with the
\kepler\ pipeline.  The target selection criteria were based upon
studying a database of 1.2$\times10^{8}$ transit injection and
recovery trials over the \kepler\ targets in order to reject targets
with noise properties making them unsuitable for an accurate
quantification of their recovery for transit signals \citep{BUR17A}.
Using this sample ensures that our estimate of $\pi({\rm SFA})$ is not
contaminated by detections around targets with ill-behaved or
overly-noisy flux time series.  The selection criteria results in
75,522 \kepler\ targets having 3900$<$\Teff$<$6000~K, \Logg$>$3.8,
similar data quantity selection as \citet{BUR15}, and excludes targets
with noise properties resulting in transit recovery statistics that do
not follow the adopted recovery model \citep{BUR17B}.  Stellar
parameters are taken from the original DR\,25 stellar catalog
\citep{MAT17}.

\begin{figure}
\hspace*{-1.5cm}\includegraphics[trim=0.0in 0.0in 0.0in 0.0in,scale=0.66,clip=true]{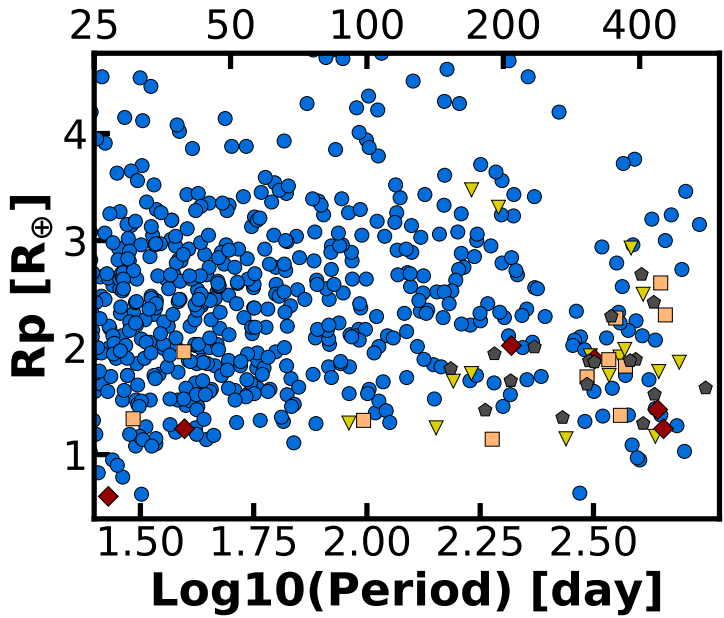}
\caption{Period versus radius for the classified planet candidates from the DR\,25 \kepler\ pipeline run (circle) and planet candidates representing systematic false alarms detected from the modified data sets using inversion (triangle), scrambling one (diamond), scrambling two (square), and scrambling three (pentagon).\label{fig:perradpc}}
\end{figure}

\begin{figure}
\hspace*{-1.0cm}\includegraphics[trim=0.0in 0.0in 0.0in 0.0in, scale=0.66,clip=true]{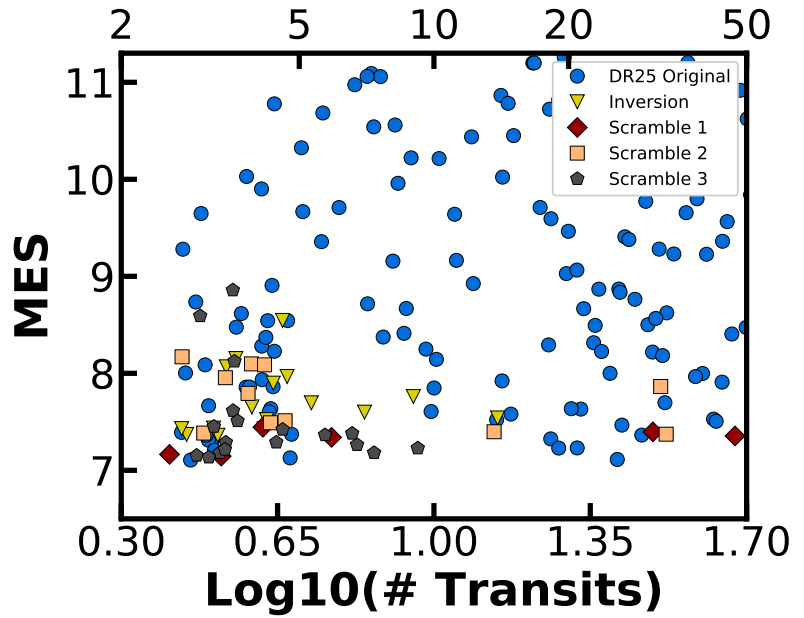}
\caption{Same data as Figure~\ref{fig:perradpc}, but in the set of parameters number of transits versus MES that are more directly related to the observational `quality' of the transit events.  The classified planet candidates from the DR\,25 \kepler\ pipeline run (circle) and planet candidates representing systematic false alarms detected from the modified data sets using inversion (triangle), scrambling one (diamond), scrambling two (square), and scrambling three (pentagon) are shown.\label{fig:ntranmespc}}
\end{figure}

We base our calculation of $\pi({\rm PL})$ on the planet candidates
identified by \citet{THO17} in the original/unmodified \kepler\ DR\,25
pipeline analysis requiring that they are associated with the above
stellar sample.  Figure~\ref{fig:perradpc} shows the distribution of
the planet candidates in \Porb\ versus \Rp\ focusing on the small,
long-period candidates (circles).  Values for \Rp\ are taken from the
DR\,25 planet candidate catalog \citep{THO17,HOF17} and are not
necessarily the best individual values available from the literature.
We base our calculation of $\pi({\rm SFA})$ on the planet candidates
identified in the four modified data sets (one from inversion and
three from scrambling) \citep{COU17A,THO17}.
Figure~\ref{fig:perradpc} shows the distribution of the planet
candidates from the modified data search and classification after our
filtering from inversion (triangle), scrambling one (diamond),
scrambling two (square), and scrambling three (pentagon).  The planet
candidates from the modified data sets concentrate toward the small,
long-period region.  However, they do extend up to \Rp$\sim$3.5 \Rear.
\Rp\ is a derived quantity, and the distribution in target stellar
radius, photometric noise, and impact parameter results in a wide
distribution of SNR values for a given \Porb\ and \Rp.  Given the
degeneracy in SNR values for a given \Porb\ and \Rp, for the remainder
of this study we will analyze the planet candidates in the more
directly measured quantities of the number of transit events detected,
\ntran\, and the SNR of the event as quantified by the MES as
calculated by the \kepler\ pipeline \citep{JEN02}.  Working in the
\ntran\ versus MES parameters improves the validity of our assumption
that $\mathcal{L}_{\rm PL}=\mathcal{L}_{\rm SFA}$ and removes the
uncertainty in the stellar parameter estimates.
Figure~\ref{fig:ntranmespc} shows the same data as in
Figure~\ref{fig:perradpc}, but with the planet candidate distribution
in \ntran\ as a function of MES.  For display purposes, a small
uniform random value is added to \ntran\ in order to avoid point
overlap as \ntran\ is an integer value.  The planet candidates from
the modified data sets, representing systematic false alarms, are
heavily concentrated near the selection threshold, MES=7.1.  {\cjb We find
that there are 16, 6, 11, and 18 planet candidates passing all the
filters from the modified data sets of inversion, scrambling one,
scrambling two, and scrambling three, respectively.  Specifically, in
addition to requiring the planet candidate to be hosted by a member of
the stellar sample, we require \Porb$>$12~day, 0.6$<$\Rp$<$6.0 \Rear,
\ntran$<100$, and MES$<$15.  There are 346 planet candidates from the
unmodified catalog that meet the same selection criteria.}
Table~\ref{tab:pcs} provides the properties for the planet candidates
shown in Figures~\ref{fig:perradpc}~and~\ref{fig:ntranmespc} from the
unmodified and modified data sets.

{\cjb There are also modified data set planet candidates extending toward
short \Porb$\sim$25 day and high \ntran$\sim$47.  We find these
targets do not have threshold crossing event (TCE) detections from the
DR\,25 \kepler\ pipeline run \citep{TWI16} and there are not KOIs
associated with these targets from the cumulative KOI table at the
NASA Exoplanet Archive.  Having false alarm contamination at lower
\Porb, higher \ntran\ was a-priori unexpected given the extremely low
probability for chance periodic alignment of systematics, emphasizing
the need to measure the contamination rate rather than rely solely on
a-priori expectations alone.  Definitively elucidating the mechanism
responsible for the short period detections is difficult since
MES$\sim$8 and \ntran$\sim$50 detections imply the individual transit
SNR$\sim$1 cannot be distinguished from noise.  From visual inspection
of the pre-search data conditioning \kepler\ light curves for these
targets, the signals appear to be due to stellar noise that was
coherent enough to trigger a detection when either inverted or
scrambled.  Human evaluation of the Robovetter diagnostic output plots
demonstrates that the human and algorithmic Robovetter struggle to
differentiate between instrumental false alarms, stellar variability,
and planet transit signals in this regime.  We examined the
possibility that the pattern of data scrambling happened to match a
bona-fide planet with transit timing variations (TTVs) that resulted
in a more uniform ephemeris with the imposed data scrambling.  The
transit events in the scrambling runs were associated with their
original timings from the unmodified data series.  We searched for
plausible orbital periods in the original unmodified data series that
result in minimum scatter in orbital phase for the transit events.  At
these preferred periods we find that the most plausible detections of
bona-fide planets with TTVs were for the SCR1 candidates with KIC
3241702 and 12453916.  The pattern of TTV requires discontinuous jumps
in the transit timings from 0.4 to 0.9 times the transit duration at
the yearly block size boundaries.  We find the SCR2 candidates with
short \Porb\ require larger 2-3 times the transit duration discrete
jumps at the scramble block boundaries to be explained by TTVs.  Given
the correlation in TTV jumps with the scrambling block boundaries, we
prefer the explanation of potential low-level stellar variability
resulting in these short period candidates from the modified data
series rather than instrumental false alarms or planet transit signals
with TTVs.}

\begin{figure}
\hspace*{-1.5cm}\includegraphics[trim=0.0in 0.0in 0.0in 0.0in, scale=0.67,clip=true]{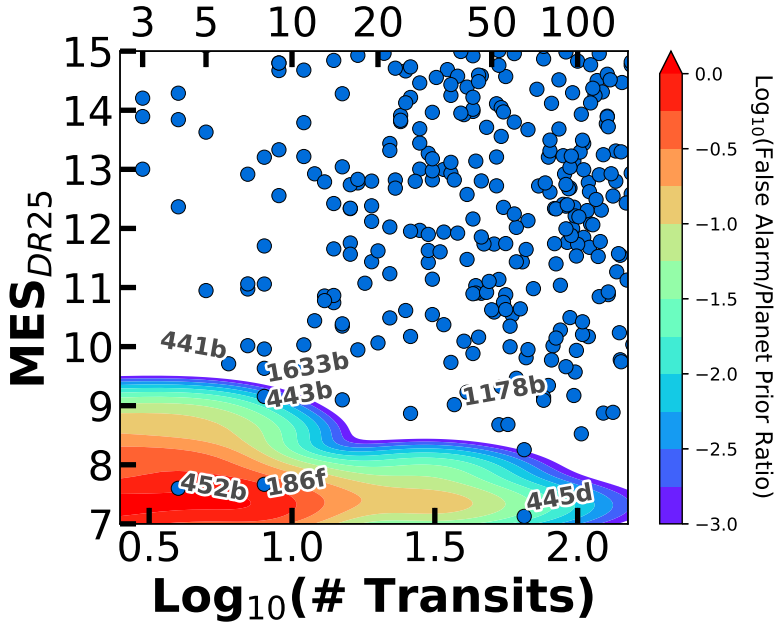}
\caption{logarithm of the odds prior ratio between the false alarm and planet hypotheses for single planet systems.  Color bar indicates best estimates of the odds ratio level.  Based upon the location of Kepler-452b (label 452b), there is a $10^{-0.07}=0.86$ odds ratio of its transit signal being due to a systematic false alarm relative to being a bona fide planet.  Multiple planet systems will have an additional reduction to the odds prior ratio (not shown in the color bar) by a factor of \mboost=10 and \mboost=14 for planets in systems with two and systems with three or more planets, respectively.\label{fig:priorrat}}
\end{figure}

Using the samples of planet candidates in Figure~\ref{fig:ntranmespc},
we calculate the number density of false alarms and planet candidates
using KDE techniques.  We apply a multivariate KDE analysis using a
Gaussian kernel with the bandwidth determine by Silverman's
rule-of-thumb that is appropriate if the true distribution is
Gaussian.  The Silverman's rule-of-thumb bases the bandwidth on the
sample standard deviation, separately in each dimension.  The KDE
analysis is performed in logarithmic in number of transits and linear
in MES.  The KDE probability densities are converted to number
densities by multiplying by the number of planet candidates in each
sample; $\pi_{\rm SFA}$ is given by the sum of the (simulated, false
alarm) planet candidates over the four modified data sets divided by
four to provide the expected number density for a single set of data.
Figure~\ref{fig:priorrat} shows the log of the relative prior
probability.  Figure~\ref{fig:priorrat}
is calculated by evaluating $\pi_{\rm r}$ at 300x300 uniform grid
spaced locations in the MES versus \ntran\ parameters.  We indicate
the specified levels of $\pi_{\rm r}$ using color shading.

Also shown in Figure~\ref{fig:priorrat} are the \kepler\ confirmed
planets (circles) using values for the DR\,25 detection MES and derived planet radius provided by the \kepler\ cumulative
table hosted by the NASA Exoplanet Archive\footnote{data retrieved
  2018, January 16}.  A select set of confirmed planets that are
discussed throughout the text are specifically labeled.  As shown by
\citet{MUL18}, we also find that Kepler-452b is in a region highly
contaminated by systematic false alarms; Kepler-186f also has
detection parameters in the systematic contaminated region.  Based on the DR\,25 data set, Kepler-452b has
  the highest $\pi_{\rm SFA}/\pi_{\rm PL}=0.86$ among the previously
  confirmed \kepler\ planets.  Using the BLENDER analysis techniques,
  \citet{JEN15} find a $\pi_{\rm AFP}/\pi_{\rm PL}=0.0023$, when
  considering astrophysical sources of false positives only, and with
  the inclusion of the false alarm hypothesis, we find that the odds
  ratio in favor of the planet hypothesis $\sim$360 times lower.
  Thus, Kepler-452b is not a statistically validated planet in agreement with \citep{MUL18}.  Besides Kepler-452b, Kepler-186f and Kepler-445d are the only other confirmed planets where the DR25 estimate of $\pi_{\rm r}$ violates the 1\% validation level.  We defer the discussion of Kepler-186f and Kepler-445d to Section~\ref{sec:multis}, where we take into account them belonging to a multiple planet system.  Other statistical
validations in the literature (in particular BLENDER) adopt a more
strict $\pi_{\rm AFP}/\pi_{\rm PL}$=1/370 or 3\,$\sigma$ confidence
in the planetary hypothesis.  Adopting this more strict threshold,  
the confirmed single planet Kepler-443b exceeds the 3\,$\sigma$
confidence in the planetary hypothesis for DR\,25.  The DR\,25 $\pi_{\rm r}$ estimates for these confirmed planets are provided in Table~\ref{tab:conf}

\section{Results for Multiple Planet Systems}\label{sec:multis}

The above discussion does not take into account the impact of
multi-planet systems.  \citet{LIS12}, \citet{LIS14}, and \citet{ROW14}
demonstrate the statistical arguments in favor of validating multiple
candidate signal systems as planets even in the face of large
($\gtrsim$0.3) fraction of false alarm rate for single detections
(such is the case for the false alarm contamination rates for
\kepler\ at the small planet, long period parameter space).
Approximately half of the previously confirmed \kepler\ planets in
Table~\ref{tab:conf} that do not meet the statistical validation
threshold are in multiple planet systems.  In this section, we extend
the arguments outlined in \citet{LIS14} to a lower SNR parameter space
containing the candidates contaminated by false alarms.

\citet{LIS14} calculate the multiplanet `boost' on the planet prior,
\mboost, that results from the planet counts and their distribution
among the targets.  The multiple planet statistics from
\citet{LIS12,LIS14} were based on half as much \kepler\ data (Q1-Q8
data) and adopted an SNR$>$10 threshold (equivalent to SNR$\gtrsim$14
for Q1-Q17 data), where SNR is the transit fit signal-to-noise ratio.
The SNR$>$10 threshold was specifically adopted to avoid false alarm
contamination as they did not have the tools to address this
contamination at the time \citep{ROW14}.  We update the single and
multiple planet system counts based upon the planet candidates
identified around our control sample of \ntarg=75,522 well behaved G
and K dwarfs observed by \kepler\ (see Section~\ref{sec:resultsing}).
The candidates under investigation in this study are well below the
thresholds of \citet{LIS14}.  We find that using the DR\,25 planet
candidate sample down to the MES=7.1 threshold (required to include
the candidates under investigation) that \mboost\ is significantly
reduced.  \citet{SIN16} also discuss variations in \mboost\ for
different planet statistics as applied to their \ktwo\ survey.

From the target control sample, the DR\,25 \kepler\ planet search and
classification finds \ncand=2125 planet candidates with \Porb$>$1.6
day \citep[same \Porb\ threshold as][]{LIS14}.  The breakdown by
planet multiplicity are 1233, 248, 85, 26, 5, and 2 targets having
one, two, three, four, five, and six planet candidates, respectively.
This yields \nmtarg=366 targets hosting two or more planet candidates.
We associate \Pgd$=1/(1+\pi_{\rm SFA}/\pi_{\rm PL})$, where \Pgd\ is
the reliability of the single candidate host sample relative to
systematic false alarms.

As described in Section~2.1 of \citet{LIS12} and further refined in
\citet{LIS14} they derive a multiplicity boost on the planet prior
based upon two methods.  The first uses the expected false alarms
relative to the observed candidates in multiplanet systems.  The
second method uses the multiple planet target counts to establish a
multiplicity boost.  Depending on the method and differences between
\citet{LIS12} and \citet{LIS14}, the multiplanet stats from
\citet{LIS14} yield a range of 16$<$\mboost$<$35 for targets with two
planets and 23$<$\mboost$<$100 for targets with three or more planets.
Using the same method, but using the multiplanet statistics from this
study, the multiplicity boost on the planet prior ranges from
10$<$\mboost$<$20 for targets with two planets and 14$<$\mboost$<$50
for targets with three or more planets.  \mboost\ is roughly half with
the full DR\,25 \kepler\ planet candidates than what is found using
the earlier sample of planet candidates at higher SNR.

The suppressed multiplicity boost is driven by the higher number of
planet candidates found with more data as the number of observed
targets is fixed.  This can be demonstrated by considering \mboost\ for
targets with three or more planets.  The dominant false positive scenario is the `two or more planets plus one false positive' case \citep[equation 6
  of][]{LIS14}.  From this contribution alone, one can show that,
\begin{equation}
  \pi_{\gamma_m}\propto\left(\frac{N_{\rm t}}{N_{\rm k}}\right)\left(\frac{N_{\rm 3+c}}{N_{\rm m}}\right),
\end{equation}
where \ntarg\ is the number of targets in the sample, \nktarg\ is the
number targets with one or more planet candidates, \nthreemcand\ is
the number of planet candidates associated with targets hosting three
or more planets, and $N_{\rm m}$ is the number of targets with two or
more planet candidates.  Using multiplanet statistics in this study,
values for \ntarg/\nktarg and \nthreemcand/\nmtarg\ are 0.58 and 0.89
times the values based upon the sample of planet candidates used in \citet{LIS14}, respectively.  Overall,
\mboost\ suppression is dominated by having more planet candidates
with more data (\ntarg/\nktarg), with a minor contribution for having
two planet systems being preferred relative to three or more planet
systems with more data (\nthreemcand/\nmtarg).

In order to account for \mboost\ in estimating the systematic false alarm contamination, we scale $\pi_{\rm r}$ by 1/\mboost.  From the DR\,25 estimate of $\pi_{\rm r}=0.54$ for Kepler-186f, equation~8 of \citet{LIS12}, and
the conservative \mboost=14 for systems with three or more candidates, Kepler-186f no longer meets the 1\% criterion for
statistical validation.  When taking into account false alarm
  contamination and \mboost, we revise the planet hypothesis
  probability of Kepler-186f to 96\%.
  \citet{QUI14} show that Kepler-186f met a 3\,$\sigma$ confidence
  ($>$99.7\% probability) for the planetary hypothesis when
  considering astrophysical false positives alone and a conservative
  \mboost\ based on the sample from \citet{LIS14} and \citet{ROW14}.
We note that Kepler-445d was not detected in the final DR\,25 \kepler\ pipeline run.  Thus, we evaluated its single star $\pi_{\rm r}$ based upon the DR\,24 \kepler\ pipeline run parameters.  However, given its DR\,24 detection MES=7.13, and the systematic MES difference between pipeline runs (see Section~\ref{sec:mesvar} and Figure~\ref{fig:mesdiff}) we consider the test of this object inconclusive and we are unable to properly determine its confirmation status relative to false alarm contamination measured in DR\,25.

\section{MES Variation}\label{sec:mesvar}

In this section, we discuss the sensitivity on determining
  the level of systematic false alarm contamination due to the
  uncertainty and systematics in determining the MES detection value
  for a given signal.  MES is approximately the SNR of the transit
signal, MES\,$\propto$\,SNR\,$\propto (\Delta/\sqrt{\sigma})\sqrt{{\rm
    N_{tran}}}$, where $\Delta$ is the transit depth and $\sigma$ is
the noise averaged over the transit duration.  However, in detail MES
is a specific quantity based upon the algorithm employed in the
\kepler\ pipeline to estimate the non-stationary noise and its
covariance.  From a single \kepler\ dataset, and considering
statistical noise alone, it is not possible to know the intrinsic
depth of a transit signal.  Analogously, from a single
\kepler\ dataset, it is not possible to know the expected MES value
since it is proportional to the intrinsic transit signal depth.
\citet{JEN02} show that for a given signal strength, the MES detection
statistic will have a Gaussian with unit variance distribution around
the intrinsic signal strength.  Thus, from the measured MES in DR\,25,
$m_{\rm DR25, \star}$, due to statistical noise, a hypothetical repeat
experiment of \kepler\ (without combining data) would result in
$p(m|m_{\rm DR25,\star})=1/\sqrt{2\pi}\exp^{-\frac{(m-m_{\rm
      DR25,\star})^2}{2}}$, where $m$ is the intrinsic MES expected
based on the intrinsic depth of the signal scaled by the noise.

\begin{figure}
\hspace*{-1.0cm}\includegraphics[trim=0.0in 0.0in 0.0in 0.0in, scale=0.66,clip=true]{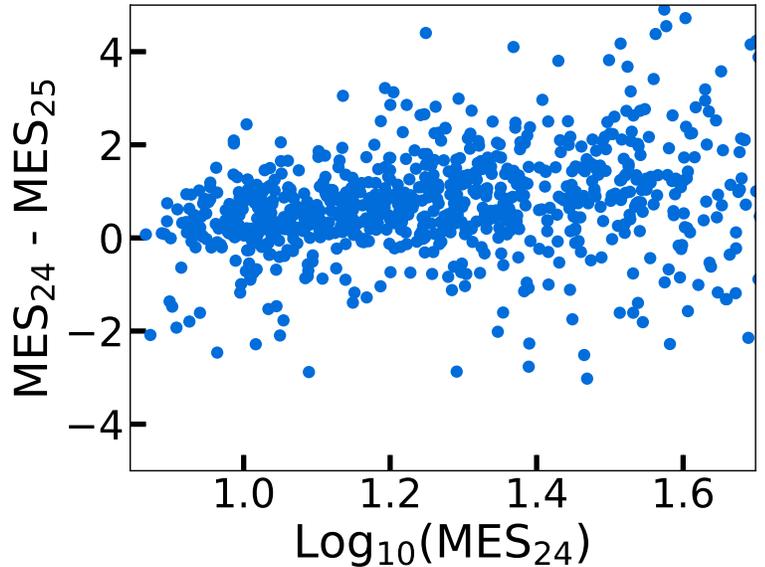}
\caption{Difference in detection MES between the final DR\,25 \kepler\ data release and the previous DR\,24 data release.  Even though the underlying data set is the same, the scatter in the detection MES difference between data releases illustrates the uncertainty in MES that needs to be taken into account when evaluating the false alarm contamination.\label{fig:mesdiff}}
\end{figure}

In addition to the intrinsic uncertainty in measuring MES,
  there is a systematic uncertainty in measuring MES.  MES is not a
  directly observed quantity.  MES is influenced by data analysis
  choices (e.g., instrument calibration, aperture selection, data
  conditioning, noise estimation, ephemeris estimate, and planet model
  fitting) that can lead to systematic variations in MES estimates.
  Thus, independent analysis of the same underlying data can lead to
  different MES estimates for a signal influenced by the numerous
  analysis choices.  We demonstrate the systematic potential for MES
variations by comparing the detection MES between the DR\,25 and the
previous \kepler\ DR\,24 release.  We show in Figure~\ref{fig:mesdiff}
the difference in detection MES between KOIs classified as planet
candidates with \ntran$<$50 that were detected in both DR\,25 and
DR\,24.  For the population of planet candidates in common with
MES$<$50, the median measured MES difference between DR\,25 and DR\,24
is $\bar{\Delta}=0.6$ with a robust mean absolute deviation of
$\sigma=0.8$.  The measured MES difference between DR\,24 and DR\,25
shows that different choices of processing algorithms can perturb the
noise components and depth estimates.  The average difference in MES
between DR\,25 and DR\,24, $\bar{\Delta}=0.6$, is in the sense that
DR\,25 has on average lower detection MES than DR\,24 for signals in
common.  The lower MES in DR\,25 primarily results from a change to
the noise estimate algorithm.  It was found that for DR\,24 and
previous releases that photometric noise estimates were systematically
overestimating the noise at the \kepler\ observing quarter boundaries
and underestimating the noise away from quarter boundaries.  The
average difference in MES between DR\,24 and DR\,25 is a warning that
comparing SNR estimates between different processing algorithms or
different pipelines and SNR calculations is fraught with systematic
biases.  Only an analysis of systematic false alarms that consistently
analyzes both detections in the observing time series and modified
time series using the same algorithms and processes can be used to
accurately assess systematic false alarm contamination.

\begin{figure}
\hspace*{-1.5cm}\includegraphics[trim=0.0in 0.0in 0.0in 0.0in, scale=0.66, clip=true]{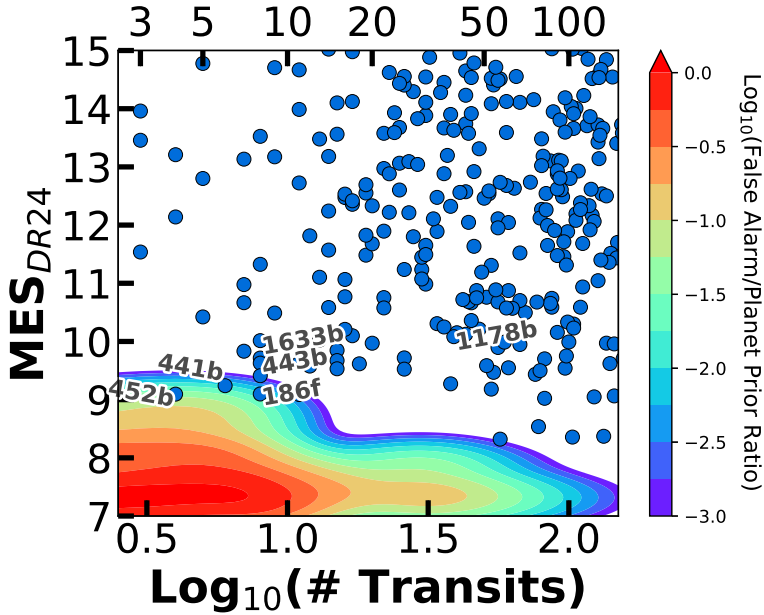}
\caption{Same as Figure~\ref{fig:priorrat}, but using the DR\,24 measured MES after removing the average difference in MES between DR24 and DR25 shown in Figure~\ref{fig:mesdiff}.  Given the level of MES variation between DR25 and DR24 and assuming a similar amplitude and pattern of the false alarm contamination for DR24 as measured in DR25, Kepler-441b would be considered an unvalidated planet.  Whereas, Kepler-186f (with the multiple planet prior boost) remain a validated planet.\label{fig:priorratdr24}}
\end{figure}

The variation in MES can have important consequences for estimating
systematic false alarm contamination between different processing
algorithms.  As an illustration, we examine a
hypothetical scenario by showing the confirmed planet locations using
their DR\,24 MES values in relation to the prior ratio estimate of DR\,25
in Figure~\ref{fig:priorratdr24}.  In order to compare to the DR\,25 prior ratio estimate,
we remove the systematic $\bar{\Delta}=0.6$ offset in MES.
Unfortunately, the data permutation data sets do not exist for the
DR\,24 \kepler\ pipeline code base, so we cannot verify that the prior
ratio estimate from DR\,25 is appropriate.  However, even though an
individual signal can have MES variations, after correcting for the
systematic MES difference, the variation is symmetrical and the prior
ratio estimate based on the full population of signals is likely to be
similar between DR\,25 and DR\,24.  Thus, we assume that $\pi_{\rm r}(m_{\rm DR24,\star})=\pi_{\rm r}(m_{\rm DR25,\star})$ after removing the systematic $\bar{\Delta}=0.6$ difference between $m_{\rm DR24,\star}$ and $m_{\rm DR25,\star}$ for signals in common.  With these assumptions, we show that using DR\,24
MES values (less $\bar{\Delta}=0.6$ MES), Kepler-452b, MES=9.09, and
Kepler-186f, MES=9.10, had higher MES estimates relative to DR\,25.
Assuming a similar prior ratio for DR\,24 as for DR\,25, Kepler-452b,
would still be unvalidated at the $>$1\% level, but including \mboost,
Kepler-186f would remain validated.  Conversely, Kepler-441b has a
lower MES in DR\,24 relative to DR\,25 and one would consider Kepler-441b
unvalidated at the $>$1\% level, where it remained validated with the
DR\,25 analysis.  Despite the contradicting validation, the MES
variations between DR\,24 and DR\,25 are within the expected MES
variations shown for the signals in common DR\,24 and DR\,25.  Due to systematic data analysis choices, independent analyses of the same underlying data would come to different conclusions regarding the validation of planet signals.

Out of concern for independent analyses to arrive at contradictory validations for an individual system, we are motivated to identify an extended set of planet confirmations where the potential for contradictory validations due to data analysis choices can occur.  To model this possibility, we calculate the predictive posterior distribution of the prior ratio in order to model a hypothetical ensemble of data reprocessings.  The predictive posterior distribution, provides the expected distribution for the prior ratio in light of a hypothetical ensemble of data reprocessings that result in an unobserved distribution of measured MES values for a given signal, $m_{\star}$, based upon the DR\,25 MES measurement, $m_{\rm DR\,25, \star}$.  The predictive posterior distribution for the prior ratio of an individual signal in an ensemble of reprocessings is given by 
\begin{equation}
p(\pi_{\rm r}|m_{\rm DR25,\star})=\int_{\Omega_{m}}p(\pi_{\rm r}|m_{\star})p(m_{\star}|m_{\rm DR25,\star})\mathop{dm_{\star}},\label{eq:post}
\end{equation}
where $m_{\star}$ is the measured MES of a hypothetical reprocessing and $\Omega_{m}$ is the integration domain over MES.  In equation~\ref{eq:post} and what follows, we have suppressed the dependence on \ntran\ as it's integral value is known without uncertainty.  We model the expected distribution of measured MES values due to reprocessing, $p(m_{\star}|m_{\rm DR25, \star})$, as a zero mean Gaussian with standard deviation, $\sigma=0.8$, based upon our empirical distribution of MES variations observed between DR\,25 and DR\,24.

In order to evaluate the conditional probability, $p(\pi_{\rm r}| m_{\star})$, we note that the prior ratio represents a direct relation between $m_{\star}$ and $\pi_{\rm r}$.  Thus, the prior ratio can be viewed as a parameter change-of-variable/probability transformation function $\phi(m_{\star})=\pi_{\rm r}(m_{\star})$.  Since a direct relationship between $m_{\star}$ and $\pi_{\rm r}$ exists, the conditional probability can be written using an expression involving the dirac generalized function, $p(\pi_{\rm r} | m_{\star})=\delta(\pi_{\rm r}-\phi(m_{\star}))$.
Furthermore, we assume that a reprocessing of the data will result in a prior ratio dependence on $m_{\star}$ that is equivalent to what was estimated in the DR\,25 analysis (i.e., $\pi_{\rm r}(m_{\star})\sim \pi_{\rm r}(m_{\rm DR25,\star})$).  This follows from the arguments given above regarding MES differences between the DR\,25 and DR\,24 processing.  The reprocessing results in a symmetric zero-mean Gaussian distributed MES variations (after removing a systematic difference in MES).
 
Due to the numerical nature of the KDE estimate for $\pi_{\rm r}$,
we calculate $p(\pi_{\rm r}|m_{\rm DR25,\star})$ through Monte-Carlo sampling of $m_{\star}$ followed by substitution into the prior ratio result.  We perform the predictive posterior distribution estimate using 25000 Monte-Carlo iterations sufficient to reach convergence of results.  In order to evaluate the KDE false alarm distribution for MES values near and below the MES=7.1 threshold, we additively reflect the false alarm distribution at MES$_{\rm pk}$=7.5 (apparent peak in the false alarm distribution) in order to extrapolate results towards lower MES. The apparent peak in the false alarm distribution is an artifact of the MES=7.1 survey threshold and the well-known bias for KDE implementations to underestimate the distribution near sharp boundaries.  In reality, the false alarm distribution continues to increase with MES$<$MES$_{\rm pk}$.  In particular, evaluation of the false alarm prior for MES=6.0, or $\delta$MES=1.5 below MES$_{\rm pk}$, involves evaluation of the KDE estimate at MES$_{\rm pk}$ and MES$_{\rm pk}$+$\delta$MES=9.0.  The logarithmic difference in the false alarm prior between MES$_{\rm pk}$ and MES$_{\rm pk}$+$\delta$MES is added to the logarithm of the false alarm prior at MES$_{\rm pk}$ in order to determine the extrapolated false alarm prior for MES$_{\rm pk}$-$\delta$MES.  The reflective extrapolation method results in a reduced skewness in the distribution for $\pi_{\rm r}$.

The resulting distribution of $\pi_{\rm r}$ for confirmed planets in and
near the false alarm contamination region spans $\gtrsim$20~dex and is
highly skew with the mode of the distribution typically occurring near
its 95$^{th}$ percentile.  In Table~\ref{tab:conf}, we report the median ($\pi_{\rm r,med}$), mode ($\pi_{\rm r,mod}$), 68.2$^{th}$ ($\pi_{\rm r,1}$), and 95$^{th}$ ($\pi_{\rm r,2}$) percentiles of the distribution of $\pi_{\rm r}$ resulting from the Monte-Carlo evaluation for a selection of previously confirmed \kepler\ planets that potentially have $\pi_{\rm r}$ distributions overlapping with a 1\% validation threshold.  We require that the mode of the distribution, $\pi_{\rm r, mod}<$0.01.  In order to ensure that the mode peak is significant relative to the broader underlying distribution, we require that the 95$^{th}$ percentile, $\pi_{\rm r,2}<$0.01.  Based upon the variations in MES observed between DR\,25 and DR\,24, it is possible that alternative analyses can influence the outcome of a strict cut on considering whether a planet remains validated.   Our model in this section considers a statistical ensemble of outcomes for influencing the measured MES values for transit signals.  If one wants to make a validation decision robust against the influence of the data analysis decisions, then this model indicates that Kepler-452b, Kepler-186f, Kepler-441b, Kepler-443b, Kepler-1633b, Kepler-1178b, and Kepler-1653b have a non-negligible probability of overlapping with the 1\% validation threshold.  {\cjbthree Given the subtle data analysis decisions that lead to a detection and a MES estimate, it is difficult to exactly assign a planet probability relative to systematic false alarms.  Furthermore, the black and white decision as to whether to consider a planet candidate as a confirmed planet or not depends upon the risk tolerance of the scientific question at hand.  In this paper, we outline a specific set of criteria that lead us to conclude that the above extended list of planets are no longer confirmed.  Other reasonable data analysis choices and criteria could lead to different black and white confirmation choices, but it should be acknowledged that including planets from the above list risks a non-negligible probability of contaminating a bona-fide planet sample at a $>$1\% level or higher as one includes detections towards lower MES and fewer \ntran.}

\section{Quietest \kepler\ Detectors}\label{sec:addanalysis}

Electronic noise properties varies between the detectors that make up the \kepler\ camera \citep{VAN16B}.  Some detectors have elevated levels of rolling band pattern noise, elevated read noise, or out-of-spec gain values.  The detectors impacted by these less ideal properties are highlighted in Table~13 of \citet{VAN16B}.  We select all the \kepler\ camera channels that have a `yellow' or `red' indicator of Table~13 in \citet{VAN16B} and exclude \kepler\ targets that pass through these noisier channels during any \kepler\ observing season.  From the 206150 \kepler\ targets that have at least one \kepler\ quarter of observations, 93079 (55\% reduction) \kepler\ targets remain after removing targets that pass through the designated noisier channels.  After the spectral type, data quality, and analysis region cuts, there are 346 observed planet candidates.  Requiring the target host to be on the quietest channels results in 150 planet candidates (57\% fewer) for analysis.  The systematic false alarm candidates are reduced from 51 down to 15 (71\% fewer).  The larger fractional reduction in false alarm candidates suggest the false alarms are slightly over-represented on the noisy channels.  Candidates and false alarms that remain for the quietest channels are indicated by a flag in Table~\ref{tab:pcs}.

\begin{figure}
\hspace*{-1.0cm}\includegraphics[trim=0.0in 0.0in 0.0in 0.0in, scale=0.67,clip=true]{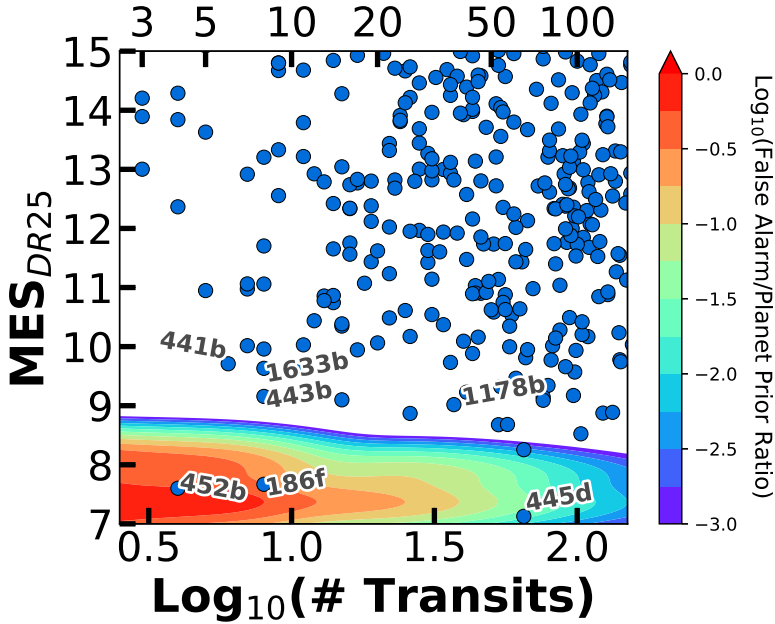}
\caption{Same as Figure~\ref{fig:priorrat}, but the detections are filtered such that only detections on the electronically `quietest'/`well-behaved' Kepler detectors are included.  For DR25 and the `well-behaved' Kepler detectors, Kepler-452b and Kepler-186f remain unvalidated due to false alarm contamination.\label{fig:priorratquiet}}
\end{figure}

Figure~\ref{fig:priorratquiet} repeats the prior ratio analysis for the subset of candidates detected on the quietest channels.  The elevated false alarm to planet prior ratio region does not extend to as high a MES at the low \ntran\ region for the quietest channel data relative to including all \kepler\ channels.  We also note that the unexpected population of false alarm candidates with \ntran$>$10 remains represented in the sample of detections on the quietest \kepler\ detectors.  For the main conclusions of this study we adopt results using targets from all channels in Section~\ref{sec:resultsing}.  We present the quiet channel analysis in order to demonstrate that the systematic false alarm contamination is present even for when limiting the analysis to the most well-behaved \kepler\ channels and is qualitatively consistent with the all channel analysis.  For DR25 and the `well-behaved' Kepler detectors, Kepler-452b and Kepler-186f remain unvalidated due to false alarm contamination.

\section{Discussion}\label{sec:discussion}

Unconfirming a planet not does imply that it is definitively a false
positive, only that the KOI should be properly considered a planet
candidate.  Even in the case of Kepler-452b, (the previously confirmed
planet suffering the highest false alarm contamination),
$P(SFA)/P(PL)$=0.86, indicating that the previously confirmed
\kepler\ planets are more likely bona-fide planets than systematic
false alarms.  The accepted convention in the literature is to
consider a planet statistically confirmed if the planet hypothesis is
favored at the 99\% level.  This threshold is arbitrary, but has the
qualitative goal of producing a sample of transiting planets confirmed
with a reliability approaching that of radial velocity surveys.
However, depending on the scientific goal and the risk posture of the
investigation, including lower fidelity planet candidates into a
confirmed planet sample may be worthwhile.

\subsection{The Case for Independent Detection}

There are several avenues to eliminate the false alarm hypothesis and
reconfirm these planets.  First, high precision radial velocity can be
obtained to detect the planet by measuring its mass using an
independent method.  A planet mass measurement is strongly aided by
knowing the orbital period of the signal a-priori rather than having
to search for the orbital period from the radial velocity data alone.
However, despite this advantage, the current radial velocity precision
and observing time available prevent radial velocity confirmation of
these faint (13$<$Kpmag$<$15) \kepler\ targets with small expected
radial velocity semiamplitude (0.3$<$\Kp$<$0.9 \mps for the planets
unconfirmed in this study).  Alternatively, the false alarm hypothesis
could be ruled out with a better model of the low-level systematics in
\kepler\ data, or an understanding of their environmental drivers, or
new vetting metrics that can more cleanly differentiate the false
alarms from planet candidates.  Machine learning techniques applied to
\kepler\ data have been helpful in this direction
\citep{MCC15,THO15,ARM17,SHA17,PEA18}, but have yet to outperform the
expert guided decision tree method employed by the Robovetter.
However, the Robovetter does employ machine learning techniques
\citep{THO15} as a subset of the decision metrics.  The additional
analysis must include data other than the flux time series alone, as
the SNR of these candidates in the flux time series is insufficient to
distinguish systematic contamination from planet transit signals.

The final possibility for eliminating the false alarm hypothesis is to
re-observe the transit events with an independent instrument.  The
long \Porb$>$ 100~day and long transit duration for these
\kepler\ candidates precludes the use of large ground-based
observatories --- the chance of a transit event optimally timed for
mid-transit near meridian crossing at large telescopes is negligible.
In addition, the ephemeris uncertainty pressure (for Kepler-452b, the
current 6 hr uncertainty in ingress time relative to mid-transit grows
$\sim$30 min every four transit events) prevents taking advantage of
possible observing chances from the ground.  \hst\ is currently the
only available resource to make timely follow-up on these important
small, long-period \kepler\ candidates.  Assuming \hst\ can achieve
orbit-to-orbit photometric stability approaching the Poisson
expectation, we expect \hst\ to achieve approximately two times higher
SNR than \kepler\ for a single transit event employing the long pass
F350LP filter on UVIS WFC3.  \spitzer\ with its smaller aperture and
relatively narrower bandpass, has a $\sim$6 times lower SNR than
\hst\ for a single transit event and G dwarf hosts.  Toward this end,
we executed a pilot \hst\ cycle 25 program (Program ID: 15129) to
recover a transit of the the habitable zone super-Earth Kepler-62f
(\Porb=267~day).  The purpose of the \hst\ program is to demonstrate
the feasibility of \hst\ to reconfirm important individual
\kepler\ discoveries.  In addition, \hst\ observations on a
statistically large sample of \kepler\ candidates is critical for an
accurate planet occurrence measurement in the regime of terrestrial,
habitable zone planets orbiting GK dwarfs in light of the significant
false alarm contamination \citep{BUR15}.  The modified data series
currently provides our best method of measuring the false alarm
contamination impacting statistical validation and planet occurrence
rate.  However, there is no guarantee that our procedure faithfully
represents the true underlying false alarms present in the original
unmodified data, and the \hst\ observations can eliminate this
concern.

\begin{figure}
\hspace*{-1.5cm}\includegraphics[trim=0.0in 0.0in 0.0in 0.0in, scale=0.67,clip=true]{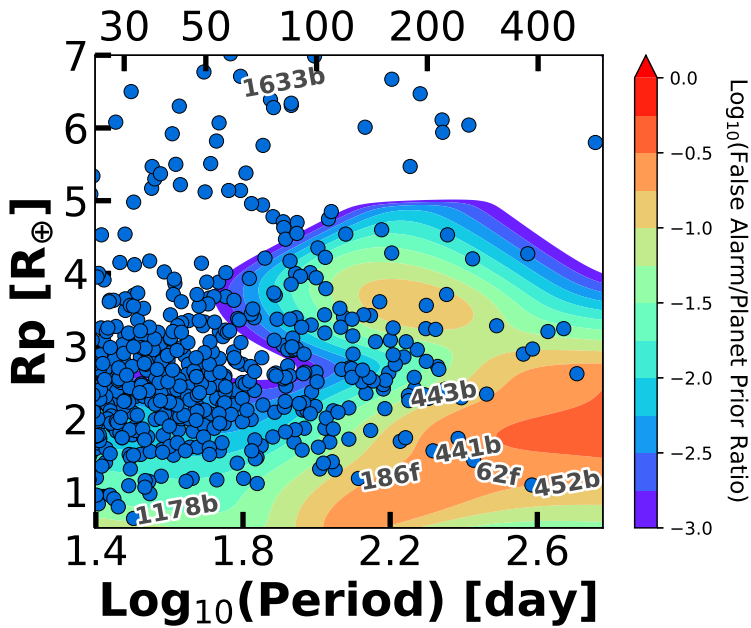}
\caption{same data as Figure~\ref{fig:priorrat}, but showing the logarithm of the odds prior ratio between the false alarm and planet hypotheses in the alternative parameters of \Porb\ and \Rp.  Color bar indicates odds ratio levels.\label{fig:priorratperrad}}
\end{figure}

Kepler-62f was chosen because when the contamination is cast in the
\Porb\ and \Rp\ parameters, it suffers enhanced false alarm
contamination.  We show in Figure~\ref{fig:priorratperrad} the same analysis as shown in Figure~\ref{fig:priorrat}, but in the alternative parameterization of \Porb\ and \Rp.
Similar to Figure~\ref{fig:perradpc}, Figure~\ref{fig:priorratperrad}
shows that due to the spread in SNR for a given location in \Porb\ and
\Rp, that an elevated systematic false alarm rate can occur over a
large region of parameter space.  From an occurrence rate perspective,
it is necessary to understand false alarms when cast in the \Porb\ and
\Rp\ basis.  However, for this study the mix of SNR levels implies a
varying likelihood which violates our assumption of $\mathcal{L}_{\rm
  PL}=\mathcal{L}_{\rm SFA}$.  Our expectation is that Kepler-62f
avoids the false alarm contamination as its MES=14.3 separates it
from the contaminating false alarms.  Kepler-62f provides a higher SNR
test case for the capabilities of \hst\ before attempting to reconfirm
lower SNR \kepler\ candidates.

\subsection{Min SNR for statistical confirmation}

It has generally been accepted in the statistical validation of
\kepler\ candidates from the prime mission \citep{ROW14,MOR16,TOR17}
that adopting SNR$>$10 for \kepler\ planet candidates is sufficient to
avoid contamination by false alarms.  As we have seen, after
quantifying an acceptable threshold to avoid false alarms, limiting
SNR$>$10 was helpful guidance.
Based on the DR\,25 analysis, from Figure~\ref{fig:priorrat}, we find that
enabling a 1\% validation relative to false alarms for
single planet systems necessitates MES$\gtrsim9.0$ for systems with \ntran$<$ 10 events and MES$\gtrsim8.0$ for 10$<$\ntran$<$60.  In
addition, we find that the MES detection statistic from the
\kepler\ pipeline is systematically lower than the SNR derived from
the transit fit.  SNR of the transit fit, rather than MES, is what
has been employed when defining a sample for validation.  For DR\,25, we
find that the median ratio between MES for detection and SNR from
the transit fit is 0.81 with a sample standard deviation of 0.5 in the
ratio.  The transit fit SNR is typically expected to be higher and can be
systematically different between various algorithms due to different
choices in pre-detection/pre-fit filter, knowing ahead of time where
event is located in the case of fitting, different choices for
duration, and depth of transit event.  Accounting for the systematic
difference between the detection MES and the transit fit SNR, the
threshold to avoid false alarms during statistical validation for
single planet systems for DR\,25 is SNR$\gtrsim$11.2 for \ntran$<$10 events and SNR$\gtrsim$10 for 10$<$\ntran$<$60 events.  Taking into account \mboost, multiple
planet systems avoid false alarms for SNR$\gtrsim10$ for \ntran$<$10
events and avoid them down to the detection threshold for \ntran$>$10.
These thresholds are appropriate for the DR\,25 pipeline analysis.  Detections from \kepler\ data using other pipelines or alternative data processing without a corresponding quantification of the false alarm population require higher thresholds.  We show in Section~\ref{sec:mesvar} that modest changes in data analysis can result in MES variations.  To be robust against potential MES variation, MES$\gtrsim10.5$ for systems with
\ntran$<$10 events and MES$\gtrsim9.5$ for 10$<$\ntran$<$60 (SNR$\gtrsim$13 and SNR$\gtrsim$11.9, respectively) is needed.  

This study provides important lessons for the continuing \ktwo\ phase
of the \kepler\ mission \citep{HOW14,VAN16} and the upcoming
\tess\ \citep{RIC16} mission.  Systematic false alarms are a dominant
contamination source preventing statistical validation when the
detection is made within several dex of the significance threshold
adopted for the survey.  To complicate the situation, the variety and
expansive set of \ktwo\ search pipelines and dissimilar data
systematics, general statements such as the above guidance based upon
\kepler\ may not apply
\citep{VAN14,ARM15,FOR15,MON15,SAN15,AIG16,CRO16,KOV16,LUG16}.  For
instance, \citet{CRO16} adopt a threshold of SNR$>$12 for detection.
Without tests, such as inversion and block bootstrap resampling, is
SNR$>$13 (like for \kepler\ single planets) sufficient to avoid false
alarm contamination, or is a $\Delta3.5$ offset in the SNR relative to
the detection threshold (implying SNR$>$15.5), or something else
entirely required?  In addition, the SNR values for the same signal
may be systematically different from one practitioner to another (as
we have demonstrated by comparing the \kepler\ MES and SNR from
transit fitting).  At least for \kepler\, we have shown that the false
alarm hypothesis exceeds the astrophysical false positive hypothesis
for the low MES, few transit event regime for the planet detections
that were previously confirmed.  The contribution of false alarms
should not be ignored when using the statistical validation method for
planet confirmation.  Furthermore, if the primary goal of a transit
study is to identify candidates for statistical validation, then
analyzing inverted and block bootstrap permuted data in order to
characterize the false alarm rate of the survey is more important than
characterizing the survey completeness through transit injection and
recovery.  However, this warning is tempered by the fact that
\ktwo\ and \tess\ missions have access to brighter targets, have a
short observing baseline $\sim$30 day, and in the case of \tess\ a
smaller aperture than for \kepler.  Thus, more traditional means of
radial velocity confirmation and ground-based transit recovery are
viable options for eliminating false alarm contamination.

The same warning applies to employing statistical validation on
samples of detections from an independent and/or `deeper' search of
\kepler\ prime mission data.  Recently, \citet{SHA17} announced
confirmation of Kepler-90i and Kepler-80g detected during an
independent search and machine learning classification of
\kepler\ prime mission data.  Candidates to Kepler-90i and Kepler-80g
where detected near a SNR$\sim$9 by \citet{SHA17}.  \citet{SHA17} did
not quantify the systematic false alarm hypothesis during the
validation of their planets.  Based upon the \kepler\ pipeline and
classification performance, we do not find significant false alarm
contamination down to the MES=7.1 threshold in the \Porb$\sim$14~day
(\ntran$\gtrsim$100) regime of these two detections.  However,
\citet{SHA17} did run their machine learning classification on the
inversion threshold crossing events (TCEs) identified by the
\kepler\ pipeline (MES$>$7.1 threshold).  Their machine learning
classifier finds four times more planet candidates than the
Robovetter.  They do not report the distribution in \Porb\ and
\Rp\ for their false alarm planet candidates, so we are unable to
determine whether they overlap with their planet confirmations.  We
conjecture that if they encounter a four times higher false alarm rate
when classifying TCEs identified by the \kepler\ pipeline employing a
MES$>7.1$ threshold, then their independent search employing a deeper
SNR$>5$ threshold will result in an insurmountable false alarm
contamination to maintain a 1\% threshold for statistical validation.
The multiple planet prior boost would also be further suppressed by
adopting a lower SNR detection threshold.

\section{Conclusion}\label{sec:conclusion}

The statistical validation technique has provided a bountiful
population of planets with which to constrain the planetary system
outcomes of planet formation.  However, we have shown that for the
\kepler\ prime mission discoveries, statistical validations currently
cannot be extended down to the detection threshold of the survey.
Most, but not all, statistical validation studies from
\kepler\ discoveries limited the planet candidate sample requiring
SNR$>$10 in the transit model fit in order to render the false alarm
contamination hypothesis negligible.  We find that this qualitative
judgment was a very helpful guide, but not conservative enough to
avoid false alarm contamination preventing $>$99\% planet probability
confirmation when faced with a quantitative measurement of the false
alarm population.  In particular, Kepler-452b and
  Kepler-186f suffer the highest level of false alarm contamination
  among the previously confirmed \kepler\ planets in DR\,25.  They have
  false alarm probabilities that exceed the 1\% validation level in
  DR\,25.  Furthermore, the extended set of planets, Kepler-441b,
  Kepler-443b, Kepler-1633b, Kepler-1653b, and Kepler-1178b have a non-negligible probability of exceeding the 1\% validation level when
  taking into account a model that considers a statistical ensemble of
  processing variations.  {\cjbthree Given the sensitivity of determining the MES for a given signal to subtle data analysis choices, it is difficult to exactly measure the planet probability relative to systematic false alarms.}  Qualitatively, the confirmation
  confidence level intends to produce a sample of transiting planets
  confirmed with a reliability approaching that of confirming the
  transit signal by additionally measuring the planet mass from radial
  velocity observations.  However, depending on the scientific goal
  and the risk posture of the investigation, including lower fidelity
  planet candidates, such as ones unconfirmed in this study, into a
  confirmed planet sample may be worthwhile. 

Our findings do not preclude the probability that the planets
unconfirmed in this study are in fact bona-fide planets.  The periodic
dimming signals representing these candidates cleanly pass a battery
of 52 vetting metrics employed by the Robovetter as consistent with a
planet transit model.  The betting odds are in favor of them being
planets.  In order to reconfirm these planet candidates, we find that
independent detection of the transit event with \hst\ represents the
most viable option for these `relatively' faint targets with
\Porb$>$100~day and shallow transits.  We estimate that \hst\ can
achieve twice the SNR per transit as \kepler.  A pilot \hst\ cycle 25
program (ID: 15129) was developed in order to demonstrate that
\hst\ can achieve the orbit-to-orbit photometric stability at the
Poisson limit in order to re-observe a transit of Kepler-62f.

Another viable option for re-confirmation is to further understand the
drivers responsible for the systematic false alarms in \kepler,
enabling their removal.  Table~\ref{tab:pcs} provides a listing of
the small, long-period candidates identified in the unmodified and
modified data sets.  A targeted examination on this short list of
candidates, in conjunction with the pixel-level transit injection database
\citep{CHR17}, may provide additional vetting metrics that can
significantly separate the false alarm detections from the
ground-truth injected signals.

Avoiding false alarm contamination for long-period \kepler\ candidates
requires selecting candidates from DR\,25 with a detection MES$>9$.  This
represents a $\Delta {\rm MES}=1.9$ above the detection threshold MES=7.1.
In addition, the systematic difference between the transit fit SNR and
detection MES results in SNR$>$11.2 equivalent threshold.  This SNR
threshold applies for single candidate systems, a SNR$>$10 threshold
is appropriate for multiple candidate systems taking into account the
multiple planet prior boost.  As guidance for other transit surveys,
such as \ktwo\ and \tess\, it is not clear whether to treat these
thresholds in an absolute sense (SNR$\gtrsim 11$) or relative sense
($\Delta$~2 dex above the survey's adopted thresholds).  In lieu of
measuring the systematic false alarm rate, we recommend adopting
  the more conservative threshold.  However, due to the relatively
brighter hosts and shorter observing baseline, many of the \ktwo\ and
\tess\ candidates can eliminate the systematic false alarm
contamination down to the survey's detection threshold by
independently confirming the transit signal with relative photometry
observations or radial velocity confirmation.

\acknowledgments

We thank Timothy Morton as the referee for insightful suggestions which
improved the manuscript.  Support for this work was provided by NASA
through grant number HST-GO-15129.010-A from the Space Telescope Science
Institute, which is operated by AURA, Inc., under NASA contract NAS
5-26555.  This research has made use of the NASA Exoplanet Archive,
which is operated by the California Institute of Technology, under
contract with the National Aeronautics and Space Administration under
the Exoplanet Exploration Program.

\software{\kepler\ Pipeline \citep{JEN17}, \kepler\ Robovetter \citep{MUL15,COU16,THO17}, NumPy \url{http://www.numpy.org}, SciPy \url{http://www.scipy.org}, Matplotlib \url{http://matplotlib.org}, statsmodels \url{http://www.statsmodels.org}}

\begin{deluxetable*}{@{\hspace{4 pt}}c|@{\hspace{4 pt}}c|@{\hspace{4 pt}}c|@{\hspace{4 pt}}c|@{\hspace{4 pt}}c|@{\hspace{4 pt}}c|@{\hspace{4 pt}}c|@{\hspace{4 pt}}c|@{\hspace{4 pt}}c}
\tablewidth{0pt}
\tablecaption{{\rm \kepler\ GK Dwarf Planet Candidate Samples}}
\tablehead{
\colhead{Data Set\tablenotemark{a}} & \colhead{KOI Number} & \colhead{KIC ID} & \colhead{Planet Number\tablenotemark{b}} & \colhead{\Porb\ \brk{day}} & \colhead{\Rp\ \brk{\Rear}}  & \colhead{MES} & \colhead{\ntran} & \colhead{Quiet Analysis}}
\startdata
OBS &  733.04 & 010271806  & 4  & 18.643 & 2.14  & 15.0  &  60 & 0 \\
OBS &  759.02 & 011018648  & 2  & 91.773 & 2.12  & 14.7  &  11 & 1 \\ 
OBS &  904.03 & 008150320  & 5  & 42.141 & 1.85  & 13.0  &  28 & 1 \\
OBS & 1338.02 & 004466677  & 2  & 42.037 & 1.66  & 12.1  &  30 & 0 \\
OBS & 1338.03 & 004466677  & 3  & 21.011 & 1.40  &  7.7  &  56 & 0 \\
OBS &  734.02 & 010272442  & 2  & 70.278 & 2.54  & 10.7  &  18 & 0 \\
OBS &  986.02 & 002854698  & 2  & 76.050 & 1.75  & 12.8  &  19 & 0 \\
OBS & 1151.04 & 008280511  & 4  & 17.453 & 0.76  &  8.4  &  73 & 0 \\
OBS &  857.02 & 006587280  & 2  & 20.026 & 2.28  &  9.3  &  54 & 1 \\
OBS & 1151.05 & 008280511  & 5  & 21.720 & 0.80  &  7.8  &  54 & 0 \\
OBS &  999.02 & 002165002  & 2  & 47.333 & 2.18  & 13.3  &  30 & 0 \\
OBS &  116.04 & 008395660  & 4  & 23.980 & 1.31  & 10.9  &  56 & 0 \\
OBS &  877.03 & 007287995  & 3  & 20.838 & 1.42  & 12.1  &  67 & 0 \\
OBS & 1031.01 & 002584163  & 1  & 14.560 & 1.64  & 12.7  &  94 & 0 \\
OBS & 1276.02 & 008804283  & 2  & 13.261 & 1.47  & 14.7  &  97 & 1 \\
OBS & 1278.04 & 008609450  & 3  & 13.640 & 1.10  & 12.0  &  96 & 1 \\
\enddata
\label{tab:pcs}
\tablenotetext{a}{OBS: Unmodified Observations; INV: Inversion; SCR1: Scrambling one; SCR2: Scrambling two; SCR3: Scrambling three}
\tablenotetext{b}{Planet number is assigned in the order that they are found in the \kepler\ pipeline search and it is used to identify the \kepler\ data products for the specific candidate.  In general, planets are identified in the pipeline in order of decreasing SNR, which can differ from increasing in \Porb\ order.}
\tablecomments{This table is available in its entirety in a machine-readable form in the online journal.  A portion is shown here for guidance regarding its form and content.}
\end{deluxetable*}

\begin{deluxetable*}{@{\hspace{1 pt}}c|@{\hspace{1 pt}}c|@{\hspace{1 pt}}c|@{\hspace{1 pt}}c|@{\hspace{1 pt}}c|@{\hspace{1 pt}}c|@{\hspace{1 pt}}c|@{\hspace{1 pt}}c|@{\hspace{1 pt}}c|@{\hspace{1 pt}}c}
\tabletypesize{\footnotesize}
\tablewidth{0pt}
\tablecaption{{\rm Confirmed \kepler\ Planets Impacted By False Alarm Contamination}}
\tablehead{
\colhead{KIC ID} & \colhead{KOI} & \colhead{Planet Name} & \colhead{$\log(\pi_{\rm r, DR25})$\tablenotemark{a}} & \colhead{$\log(\pi_{\rm r, med})$\tablenotemark{b}} & \colhead{$\log(\pi_{\rm r, mod})$\tablenotemark{c}} & \colhead{$\log(\pi_{\rm r,1})$\tablenotemark{d}} & \colhead{$\log(\pi_{\rm r,2})$\tablenotemark{e}} & \colhead{\ntran}  & \colhead{MES}} 
\startdata
8311864 & 7016.01 & Kepler-452b  & -0.07 & -0.06  & -0.10 &  0.24 &  0.96 & 4  & 7.60 \\
8120608 & 571.05 & Kepler-186f   & -0.27 & -0.28  & -0.15 &  0.09 &  1.1 & 8  & 7.67 \\
9730163 & 2704.03 & Kepler-445d  & -1.60 & -0.91  & -1.11 & -0.03 &  7.41 & 65 & 7.13 \\
11757451 & 4745.01 & Kepler-443b & -2.07 & -2.24  & -1.19 & -1.44 & -0.48 & 8  & 9.16 \\
11284772 & 4622.01 & Kepler-441b & -4.87 & -4.73  & -1.14 & -2.42 & -0.86 & 6  & 9.71 \\
8745553 & 5568.01 & Kepler-1633b & -4.90 & -4.64  & -1.24 & -2.56 & -1.16 & 8  & 9.63 \\
8950568 & 2038.04 & Kepler-85e   & -5.04 & -5.06  & -1.95 & -2.86 & -1.07 & 53 & 8.68 \\
6026438 & 2045.03 & Kepler-354c  & -5.20 & -5.21  & -1.92 & -3.00 & -1.15 & 57 & 8.68 \\
5956342 & 1052.04 & Kepler-265d  & -5.90 & -5.48  & -1.24 & -3.29 & -0.94 & 26 & 8.87 \\
5977470 & 4550.01 & Kepler-1653b & -5.93 & -5.53  & -1.94 & -3.39 & -1.82 & 11 & 9.64 \\
10858691 & 1306.04 & Kepler-286e & -7.61 & -7.51  & -1.41 & -4.31 & -1.22 & 37 & 9.02 \\
6021275 & 284.04 & Kepler-132e   & -7.70 & -7.45  & -1.33 & -4.26 & -1.31 & 8  & 9.96 \\
9896558 & 1718.02 & Kepler-937c  & -7.96 & -7.77  & -1.15 & -4.39 & -1.18 & 7  & 10.02\\
9334893 & 2298.02 & Kepler-1178b & -9.61 & -9.53  & -1.61 & -6.07 & -1.54 & 41 & 9.23 \\
\enddata
\label{tab:conf}
\tablenotetext{a}{DR25 estimate of the prior odds ratio between the systematic false alarm scenario and planet scenario, $\pi_{\rm r}$.}
\tablenotetext{b}{Median estimate of the $\pi_{\rm r}$ distribution using a model accounting for processing variations.}
\tablenotetext{c}{Mode estimate of the $\pi_{\rm r}$ distribution using a model accounting for processing variations.}
\tablenotetext{d}{68.2$^{th}$ percentile (1-$\sigma$) estimate of the $\pi_{\rm r}$ distribution using a model accounting for processing variations.}
\tablenotetext{e}{95$^{th}$ percentile (2-$\sigma$) estimate of the $\pi_{\rm r}$ distribution using a model accounting for processing variations.} 
\end{deluxetable*}


\begin{thebibliography}

\bibitem[Aigrain et al.(2016)]{AIG16} Aigrain, S., Parviainen, H., \& Pope, B.~J.~S.\ 2016, \mnras, 459, 2408

\bibitem[Armstrong et al.(2015)]{ARM15} Armstrong, D.~J., Kirk, J., Lam, K.~W.~F., et al.\ 2015, \aap, 579, A19 

\bibitem[Armstrong et al.(2017)]{ARM17} Armstrong, D.~J., Pollacco, D., \& Santerne, A.\ 2017, \mnras, 465, 2634 

\bibitem[Batalha et al.(2013)]{BAT13} Batalha, N.~M., Rowe, J.~F.,
  Bryson, S.~T., et al.\ 2013, \apjs, 204, 24

\bibitem[Borucki et al.(2011a)]{BOR11A} Borucki, W.~J., Koch, D.~G., Basri, G., et al.\ 2011a, \apj, 728, 117 

\bibitem[Borucki et al.(2011b)]{BOR11B} Borucki, W.~J., Koch, D.~G., Basri, G., et al.\ 2011b, \apj, 736, 19 

\bibitem[Brown(2003)]{BRO03} Brown, T.~M.\ 2003, \apjl, 593, L125 

\bibitem[Bryson et al.(2017)]{BRY17} Bryson, S.~T., Abdul-Masih, M., Batalha, N., et al.\ 2017, Kepler Science Document, KSCI-19093-003, Edited by Michael R.~Haas and Natalie M.~Batalha, 

\bibitem[Burke et al.(2014)]{BUR14} Burke, C.~J., Bryson, S.~T., Mullally, F., et al.\ 2014, \apjs, 210, 19

\bibitem[Burke et al.(2015)]{BUR15} Burke, C.~J., Christiansen, J.~L., Mullally, F., et al.\ 2015, \apj, 809, 8 
  
\bibitem[Burke \& Catanzarite(2017a)]{BUR17A} Burke, C.~J., \& Catanzarite, J.\ 2017, Kepler Science Document, KSCI-19109-002, Edited by Michael R.~Haas and Natalie M.~Batalha,  
  
\bibitem[Burke \& Catanzarite(2017b)]{BUR17B} Burke, C.~J., \& Catanzarite, J.\ 2017, Kepler Science Document, KSCI-19111-002, Edited by Michael R.~Haas and Natalie M.~Batalha

\bibitem[Caldwell et al.(2010)]{CAL10} Caldwell, D.~A., Kolodziejczak, J.~J., Van Cleve, J.~E., et al.\ 2010, \apjl, 713, L92 

\bibitem[Christiansen et al.(2012)]{CHR12} Christiansen, J.~L., Jenkins, J.~M., Caldwell, D.~A., et al.\ 2012, \pasp, 124, 1279 

\bibitem[Christiansen(2017)]{CHR17} Christiansen, J.~L.\ 2017, Kepler Science Document, KSCI-19110-001, Edited by Michael R.~Haas and Natalie M.~Batalha
    
\bibitem[Coughlin et al.(2016)]{COU16} Coughlin, J.~L., Mullally, F., Thompson, S.~E., et al.\ 2016, \apjs, 224, 12 


\bibitem[Coughlin(2017a)]{COU17A} Coughlin, J.~L.\ 2017a, Kepler Science Document, KSCI-19105-002, Edited by Natalie Batalha and Michael R.~Haas  

\bibitem[Coughlin(2017b)]{COU17B} Coughlin, J.~L.\ 2017b, Kepler Science Document, KSCI-19114-002, Edited by Natalie Batalha and Michael R.~Haas,

\bibitem[Crossfield et al.(2016)]{CRO16} Crossfield, I.~J.~M., Ciardi, D.~R., Petigura, E.~A., et al.\ 2016, \apjs, 226, 7 

\bibitem[D{\'{\i}}az et al.(2014)]{DIA14} D{\'{\i}}az, R.~F., Almenara, J.~M., Santerne, A., et al.\ 2014, \mnras, 441, 983

\bibitem[Foreman-Mackey et al.(2015)]{FOR15} Foreman-Mackey, D., Montet, B.~T., Hogg, D.~W., et al.\ 2015, \apj, 806, 215 

  
\bibitem[Fressin et al.(2011)]{FRE11} Fressin, F., Torres, G., D{\'e}sert, J.-M., et al.\ 2011, \apjs, 197, 5

\bibitem[Hoffman \& Rowe(2017)]{HOF17} Hoffman, K., L., \& Rowe, J.~F.\ 2017, Kepler Science Document, KSCI-19113-001, Edited by Michael R.~Haas and Natalie M.~Batalha

\bibitem[Howell et al.(2014)]{HOW14} Howell, S.~B., Sobeck, C., Haas, M., et al.\ 2014, \pasp, 126, 398 

  
\bibitem[Jenkins et al.(2002)]{JEN02} Jenkins, J.~M., Caldwell, D.~A., \& Borucki, W.~J.\ 2002, \apj, 564, 495  

\bibitem[Jenkins et al.(2010)]{JEN10} Jenkins, J.~M., Caldwell, D.~A., Chandrasekaran, H., et al.\ 2010, \apjl, 713, L87
  
\bibitem[Jenkins et al.(2015)]{JEN15} Jenkins, J.~M., Twicken, J.~D., Batalha, N.~M., et al.\ 2015, \aj, 150, 56
  
\bibitem[Jenkins(2017)]{JEN17} Jenkins, J.~M.\ 2017, Kepler Science Document, KSCI-19081-002, Edited by Jon M.~Jenkins.

\bibitem[Kolodziejczak et al.(2010)]{KOL10} Kolodziejczak, J.~J., Caldwell, D.~A., Van Cleve, J.~E., et al.\ 2010, \procspie, 7742, 77421G 

\bibitem[Kov{\'a}cs et al.(2016)]{KOV16} Kov{\'a}cs, G., Hartman, J.~D., \& Bakos, G.~{\'A}.\ 2016, \aap, 585, A57 

\bibitem[Kruse \& Agol(2014)]{KRU14} Kruse, E., \& Agol, E.\ 2014, Science, 344, 275 
  
\bibitem[Lissauer et al.(2012)]{LIS12} Lissauer, J.~J., Marcy, G.~W., Rowe, J.~F., et al.\ 2012, \apj, 750, 112 

\bibitem[Lissauer et al.(2014)]{LIS14} Lissauer, J.~J., Marcy, G.~W., Bryson, S.~T., et al.\ 2014, \apj, 784, 44 

\bibitem[Luger et al.(2016)]{LUG16} Luger, R., Agol, E., Kruse, E., et al.\ 2016, \aj, 152, 100 

\bibitem[Mandushev et al.(2005)]{MAN05} Mandushev, G., Torres, G., Latham, D.~W., et al.\ 2005, \apj, 621, 1061 


\bibitem[Torres et al.(2004)]{TOR04} Torres, G., Konacki, M., Sasselov, D.~D., \& Jha, S.\ 2004, \apj, 614, 979 


\bibitem[Mathur et al.(2017)]{MAT17} Mathur, S., Huber, D., Batalha, N.~M., et al.\ 2017, \apjs, 229, 30
  
\bibitem[McCauliff et al.(2015)]{MCC15} McCauliff, S.~D., Jenkins, J.~M., Catanzarite, J., et al.\ 2015, \apj, 806, 6 

\bibitem[Montet et al.(2015)]{MON15} Montet, B.~T., Morton, T.~D., Foreman-Mackey, D., et al.\ 2015, \apj, 809, 25 


\bibitem[Morton \& Johnson(2011)]{MOR11} Morton, T.~D., \& Johnson, J.~A.\ 2011, \apj, 738, 170 

\bibitem[Morton(2012)]{MOR12} Morton, T.~D.\ 2012, \apj, 761, 6 

\bibitem[Morton et al.(2016)]{MOR16} Morton, T.~D., Bryson, S.~T., Coughlin, J.~L., et al.\ 2016, \apj, 822, 86 
  
\bibitem[Mullally et al.(2015)]{MUL15} Mullally, F., Coughlin, J.~L., Thompson, S.~E., et al.\ 2015, \apjs, 217, 31

\bibitem[Mullally et al.(2016)]{MUL16} Mullally, F., Coughlin, J.~L., Thompson, S.~E., et al.\ 2016, \pasp, 128, 074502

\bibitem[Mullally(2017)]{MUL17} Mullally, F.\ 2017, Kepler Science Document, KSCI-19115-001, Edited by Natalie Batalha and Michael R.~Haas  

\bibitem[Mullally et al.(2018)]{MUL18} Mullally, F. et al.\ 2018, \aj, in press

\bibitem[Quintana et al.(2014)]{QUI14} Quintana, E.~V., Barclay, T., Raymond, S.~N., et al.\ 2014, Science, 344, 277 

\bibitem[Ricker et al.(2016)]{RIC16} Ricker, G.~R., Vanderspek, R., Winn, J., et al.\ 2016, \procspie, 9904, 99042B 

\bibitem[Rowe et al.(2014)]{ROW14} Rowe, J.~F., Bryson, S.~T., Marcy, G.~W., et al.\ 2014, \apj, 784, 45 

\bibitem[Rowe et al.(2015)]{ROW15} Rowe, J.~F., Coughlin, J.~L., Antoci, V., et al.\ 2015, \apjs, 217, 16

\bibitem[Sanchis-Ojeda et al.(2015)]{SAN15} Sanchis-Ojeda, R., Rappaport, S., Pall{\`e}, E., et al.\ 2015, \apj, 812, 112 

\bibitem[Pearson et al.(2018)]{PEA18} Pearson, K.~A., Palafox, L., \& Griffith, C.~A.\ 2018, \mnras, 474, 478 

\bibitem[Sahu \& Gilliland(2003)]{SAH03} Sahu, K.~C., \& Gilliland, R.~L.\ 2003, \apj, 584, 1042 

\bibitem[Shallue \& Vanderburg(2018)]{SHA17} Shallue, C.~J., \& Vanderburg, A.\ 2018, \aj, 155, 94 


\bibitem[Sinukoff et al.(2016)]{SIN16} Sinukoff, E., Howard, A.~W., Petigura, E.~A., et al.\ 2016, \apj, 827, 78 


\bibitem[Stumpe et al.(2012)]{STU12} Stumpe, M.~C., Smith, J.~C., Van Cleve, J.~E., et al.\ 2012, \pasp, 124, 985

\bibitem[Thompson et al.(2012)]{THO12} Thompson, S.~E., Everett, M., Mullally, F., et al.\ 2012, \apj, 753, 86

\bibitem[Thompson et al.(2015)]{THO15} Thompson, S.~E., Mullally, F., Coughlin, J., et al.\ 2015, \apj, 812, 46 

\bibitem[Thompson et al.(2018)]{THO17} Thompson, S.~E., Coughlin, J.~L., Hoffman, K., et al.\ 2018, \apjs, 235, 38 

\bibitem[Torres et al.(2011)]{TOR11} Torres, G., Fressin, F., Batalha, N.~M., et al.\ 2011, \apj, 727, 24

\bibitem[Torres et al.(2015)]{TOR15} Torres, G., Kipping, D.~M., Fressin, F., et al.\ 2015, \apj, 800, 99 

\bibitem[Torres et al.(2017)]{TOR17} Torres, G., Kane, S.~R., Rowe, J.~F., et al.\ 2017, \aj, 154, 264 

\bibitem[Twicken et al.(2016)]{TWI16} Twicken, J.~D., Jenkins, J.~M., Seader, S.~E., et al.\ 2016, \aj, 152, 158

\bibitem[Van Cleve et al.(2016)]{VAN16} Van Cleve, J.~E., Howell, S.~B., Smith, J.~C., et al.\ 2016, \pasp, 128, 075002 

\bibitem[Van Cleve \& Caldwell(2016)]{VAN16B} Van Cleve, J.~E., \& Caldwell, D.~A.\ 2016, Kepler Science Document, KSCI-19033-002, Edited by Michael R.~Haas and Steve B.~Howell

\bibitem[Vanderburg \& Johnson(2014)]{VAN14} Vanderburg, A., \& Johnson, J.~A.\ 2014, \pasp, 126, 948 


\end{thebibliography}
\end{document}